\input{aipcheck}

\documentclass[final
    ,final            % use final for the camera ready runs
  ]
  {aipproc}

\usepackage{bm}

\layoutstyle{6x9}

\begin{document}

\title{Statics and dynamics of a harmonic oscillator coupled to a one-dimensional Ising system}


\author{A. Prados}{
  address={F\'{\i}sica Te\'{o}rica, Univ. de Sevilla,
Apartado de Correos 1065, 41080 Sevilla, Spain}
}

\author{L. L. Bonilla}{
  address={G. Mill\'an Institute for Fluid Dynamics, Nanoscience and Industrial Mathematics, Univ. Carlos III de Madrid, 28911 Legan\'es, Spain}
}

\author{A. Carpio}{
  address={Departamento de Matem\'atica Aplicada, Univ. Complutense de Madrid, 28040 Madrid, Spain} % additional visiting address
}

\begin{abstract}
We investigate an oscillator linearly coupled with a one-dimensional Ising system.  The coupling gives rise to drastic changes both in the oscillator statics and dynamics. Firstly, there appears a second order phase transition, with the  oscillator stable rest position as its order parameter. Secondly, for fast spins, the oscillator dynamics is described by an effective equation with a nonlinear friction term that drives the oscillator towards the stable equilibrium state.
\end{abstract}

\maketitle

%%%%%%%%%%%%%%%%%%%%%%%%%%%%%%%%%%%%%%%%%%%%
%% MAINMATTER
%%%%%%%%%%%%%%%%%%%%%%%%%%%%%%%%%%%%%%%%%%%%

We analise the behaviour of a classical harmonic oscillator (mass $m$, frequency $\omega_0$, position $x$ and momentum $p$) linearly coupled to a one dimensional Ising system ($N$ spin variables $\sigma_i=\pm 1$) in contact with a heat bath at temperature $T$. After a suitable nondimensionalization of the variables, the system energy per spin site is \cite{PByC10}
\begin{equation}\label{1}
 {\cal H}(x,p,\bm{\sigma})  =  \frac{p^2+x^2}{2}-\frac{x}{N}\sum_{i=1}^N \sigma_i \sigma_{i+1} \,,
\end{equation}
The oscillator dynamics follows from Hamilton's equations,
%\begin{equation}\label{2}
$\ddot{x}+x=(1/N)\sum_{i=1}^N \sigma_i \sigma_{i+1}$,
%\end{equation}
while the spins dynamics is governed by a master equation with Glauber single-flip transition rates \cite{Gl63}. Then,
%\begin{equation}\label{3}
%\end{equation}
the rate at which the $i$-th spin flips in configuration $\bm{\sigma}$ is $W_i(\bm{\sigma}|x,p)=\delta \left[
1-\gamma \sigma_{i} (\sigma_{i-1}+\sigma_{i+1})/2 \right]/2$, with
$\gamma=\tanh(2 x/ \theta)$. In these rates, $\delta$ gives the (dimensionless) characteristic attempt rate for the spin flips and $\theta$ is the dimensionless temperature \cite{PByC10}. The equilibrium properties of the system are obtained from the canonical distribution
%\begin{equation}\label{4}
${\cal P}_{{eq}}(x,p,\bm{\sigma})\propto\exp\left[-N\,
    {\cal H}(x,p,\bm{\sigma})/\theta \right].$
%\end{equation}
By summing over the spin variables, we calculate the marginal probability for the oscillator variables,
\begin{equation}\label{5}
{\cal P}_{{eq}}(x,p)\propto \exp \left[-N \left(\frac{p^2}{2}
   +\mathcal{V}_{{eff}}(x)\right)/\theta \right]  \,, \quad
    \mathcal{V}_{{eff}}(x)= \frac{x^2}{2}- \theta\left[\ln\cosh\left(\frac{x}{\theta}\right)+\ln 2\right]
\end{equation}
in which the spins produce a non-harmonic additional contribution to the effective potential of  the oscillator, $\mathcal{V}_{{eff}}(x)$. The equilibrium stable rest points $\widetilde{x}_{{eq}}$ of the oscillator are the minima of $\mathcal{V}_{{eff}}$, so that they verify the bifurcation equation
%\begin{equation}\label{6}
  $\widetilde{x}_{{eq}}- \tanh(\widetilde{x}_{{eq}}/\theta)=0$.
%\end{equation}
Thus, there is a second order phase transition at $\theta=1$, with $\widetilde{x}_{{eq}}$ as its order parameter.% For $\theta>1$, $\widetilde{x}_{{eq}}=0$, while $\widetilde{x}_{{eq}}$ separates from zero for $\theta<1$, $\widetilde{x}_{{eq}}\sim\sqrt{3(1-\theta)}$ in the limit as $\theta\to 1^-$.

\begin{ltxfigure}
\begin{center}
\includegraphics[scale=0.25]{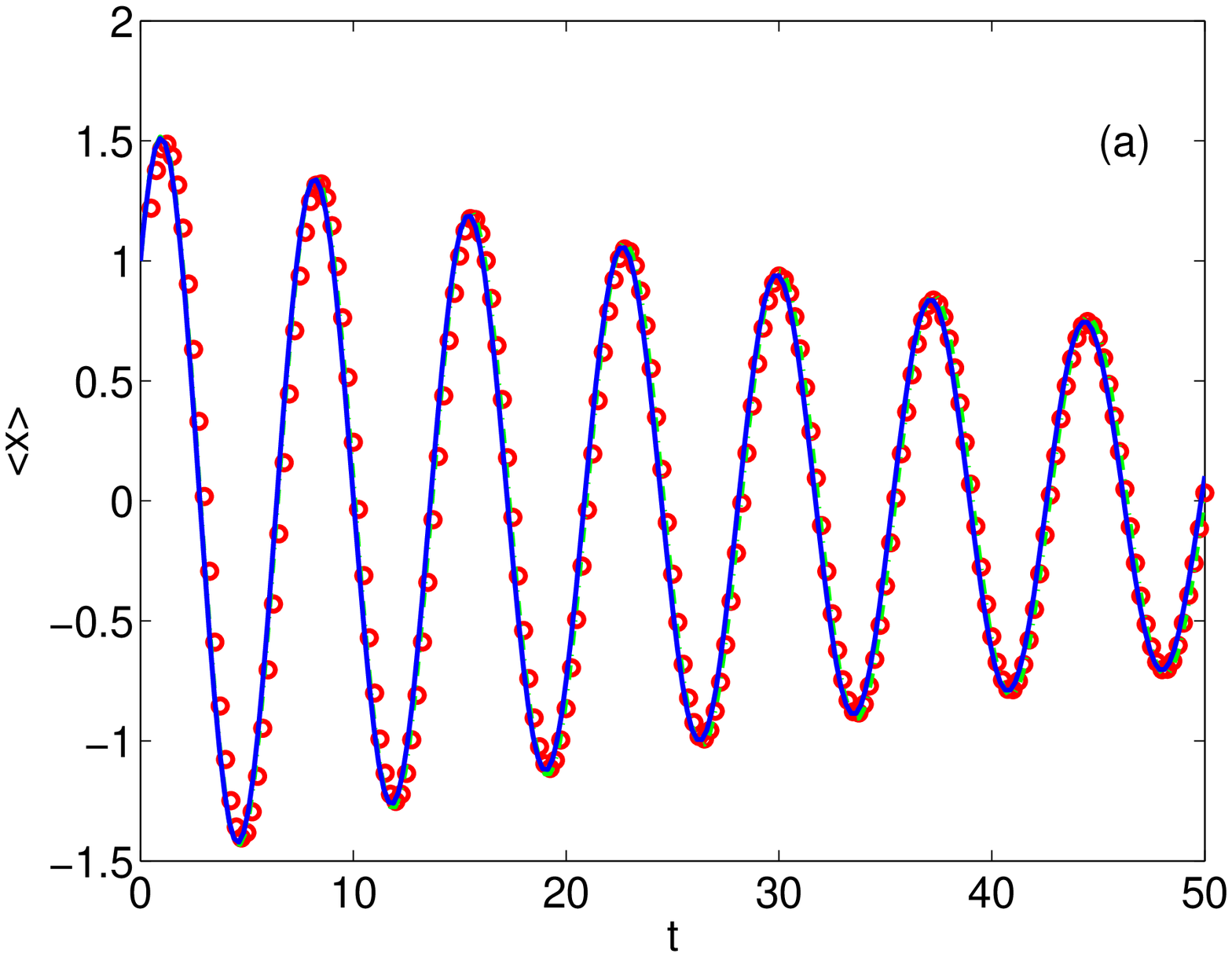}
\includegraphics[scale=0.25]{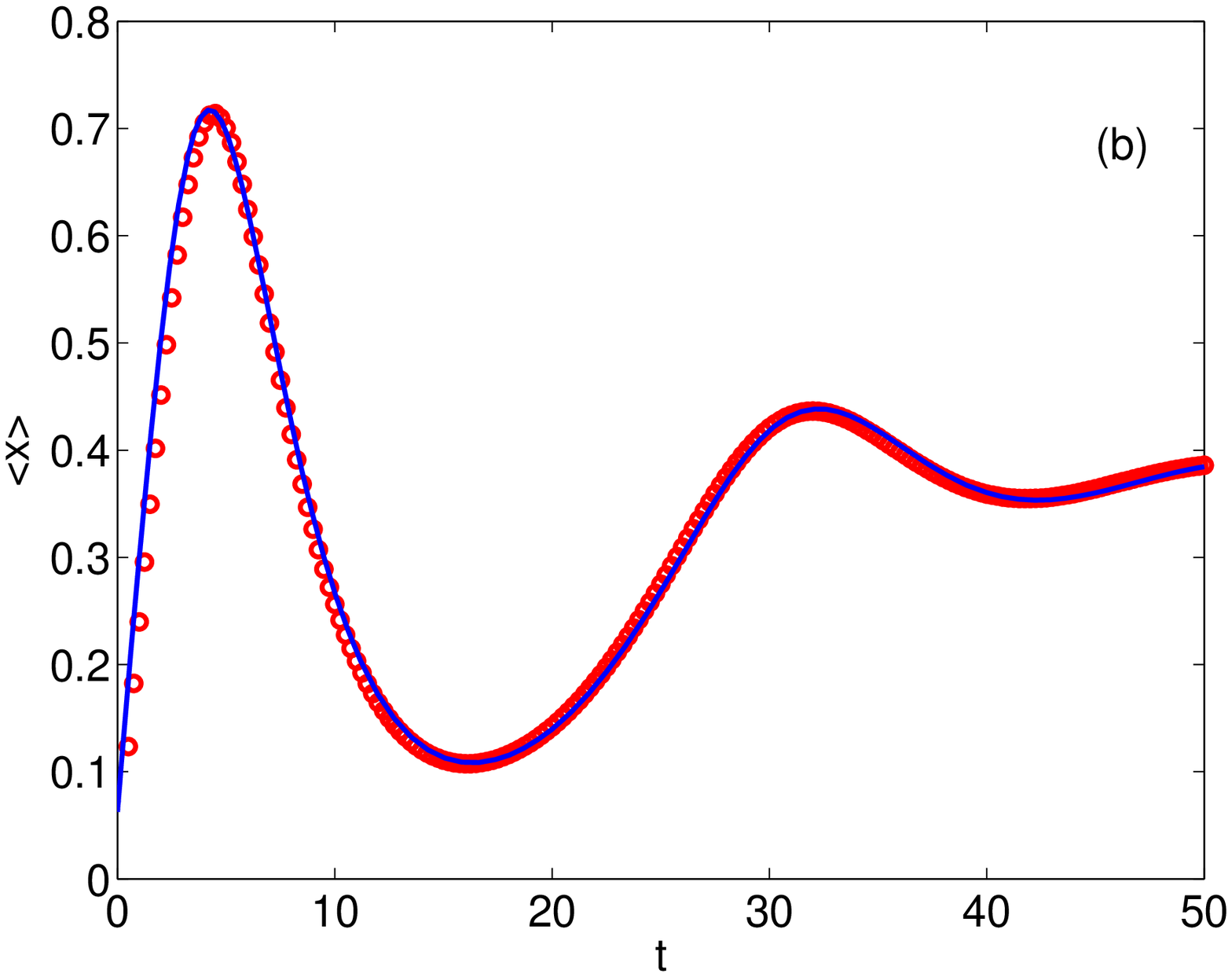} \includegraphics[scale=0.25]{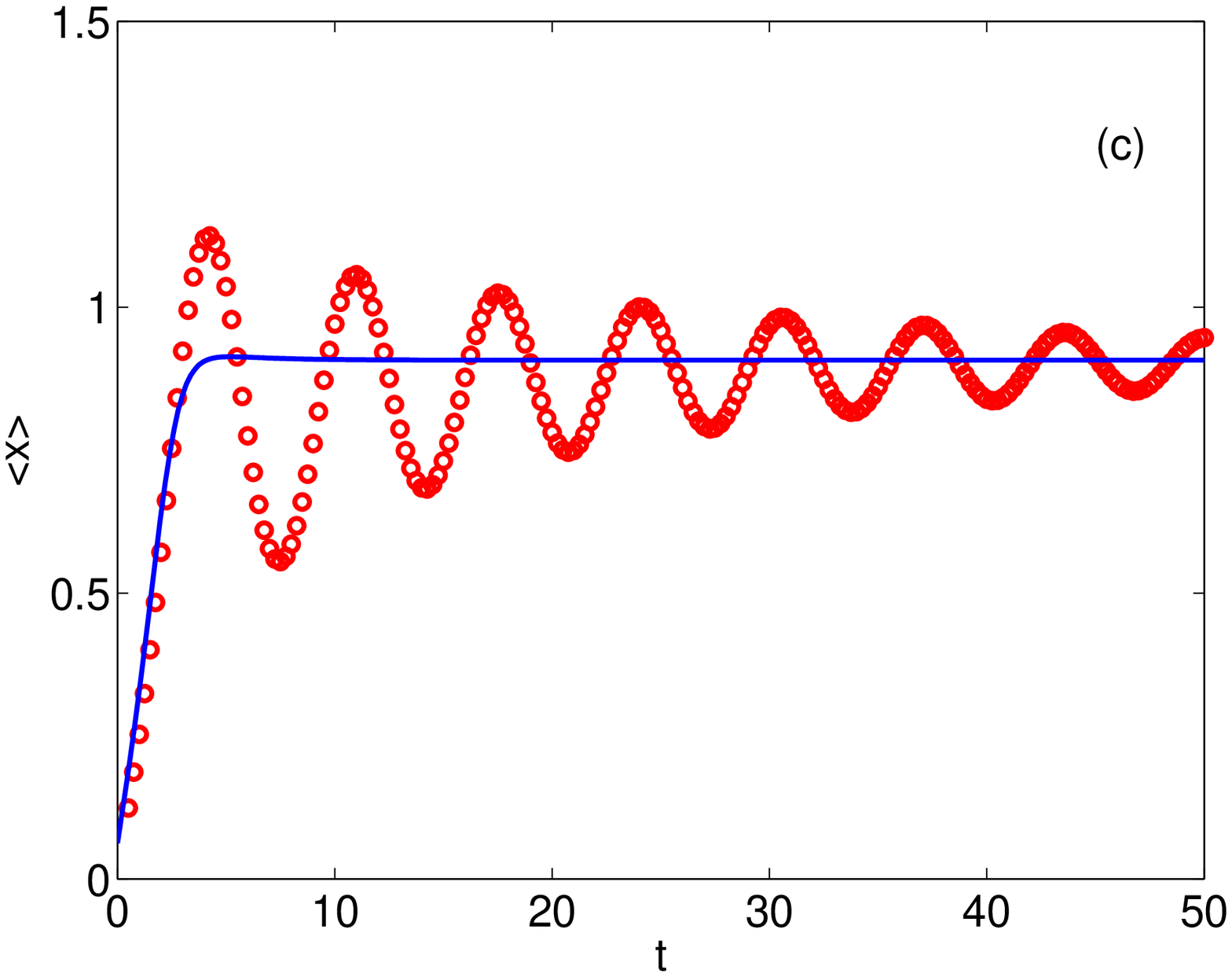}
\end{center}
\caption{Averaged stochastic trajectories $\langle x(t)\rangle=\widetilde{x}(t)$ (circles) versus nonlinear (solid blue line) predictions for (a) $\theta=4$, (b) $\theta=0.95$, and (c) $\theta=0.6$.}
\end{ltxfigure}

In the limit of fast spins, a perturbative analysis in powers of $\delta^{-1}\ll 1$ yields an approximate nonlinear equation for the macroscopic (completely neglecting the fluctuations) value of the oscillator position $\widetilde{x}$
\begin{equation}\label{7}
    \frac{d^2\widetilde{x}}{d t^{2}}=- \mathcal{V}^{\prime}_{{eff}}(
\widetilde{x})- \delta^{-1}\frac{1}{2\theta}R(\widetilde{x})\, \frac{d\widetilde{x}}{d t}, \qquad R(\widetilde{x}) =\frac{1+\tanh^2( \frac{\widetilde{x}}{\theta})}{1-\tanh^2( \frac{\widetilde{x}}{\theta})} \,.
\end{equation}
The spins are so fast that they relax, almost instantaneously, to the equilibrium state corresponding to $\widetilde{x}(t)$. But, since $\widetilde{x}(t)$ is slowly varying, there is a small separation from equilibrium proportional to $d\widetilde{x}/dt$, which gives rise to the friction term.

In the figure, the theoretical result (\ref{7}) (its numerical integration) is compared to the simulation of the stochastic model, for a random initial configuration of the spins and suitable initial conditions for the oscillator \cite{PByC10}. For high temperatures, case (a), and  for a temperature below (but close to) the critical value $\theta=1$,  case (b), the agreement between simulation and theory is excellent. On the other hand, for a further lower value of the temperature, case (c), the theory breaks down. This was to be expected, since for very low temperatures the spins relaxation time diverge and they are actually slow as compared with the oscillator. Hence, another approach is needed, in which the fast vibrations of the oscillator are averaged using the method of multiple scales \cite{BPyC10}.

%%%%%%%%%%%%%%%%%%%%%%%%%%%%%%%%%%%%%%%%%%%%
%% Sample figure:
%%
%% The option [height=...] scales the picture to the given height,
%% without it it would be printed at its nominal size
%%%%%%%%%%%%%%%%%%%%%%%%%%%%%%%%%%%%%%%%%%%%

%%%%%%%%%%%%%%%%%%%%%%%%%%%%%%%%%%%%%%%%%%%%%%%%
%% BACKMATTER
%%%%%%%%%%%%%%%%%%%%%%%%%%%%%%%%%%%%%%%%%%%%%%%%

\begin{theacknowledgments}
  This research has been supported by the Spanish Ministerio de Ciencia e Innovaci\'on (MICINN) through Grants No. FIS2008-01339 (AP), FIS2008-04921-C02-01 (LLB), and FIS2008-04921-C02-02 (AC). We would like also to thank the Spanish National Network Physics of Out-of-Equilibrium Systems (MICINN Grant FIS2008-04403-E).

\end{theacknowledgments}

%%%%%%%%%%%%%%%%%%%%%%%%%%%%%%%%%%%%%%%%%%%%%%%%
%% The bibliography can be prepared using the BibTeX program or
%% manually.
%%
%% The code below assumes that BibTeX is used.  If the bibliography is
%% produced without BibTeX comment out the following lines and see the
%% aipguide.pdf for further information.
%%
%% For your convenience a manually coded example is appended
%% after the \end{document}
%%%%%%%%%%%%%%%%%%%%%%%%%%%%%%%%%%%%%%%%%%%%%%%%

%%%%%%%%%%%%%%%%%%%%%%%%%%%%%%%%%%%%%%%%%%%%%%%%
%% You may have to change the BibTeX style below, depending on your
%% setup or preferences.
%%
%%
%% For The AIP proceedings layouts use either
%%%%%%%%%%%%%%%%%%%%%%%%%%%%%%%%%%%%%%%%%%%%

\bibliographystyle{aipproc}   % if natbib is available
%\bibliographystyle{aipprocl} % if natbib is missing

%%%%%%%%%%%%%%%%%%%%%%%%%%%%%%%%%%%%%%%%%%%
%% You probably want to use your own bibtex database here
%%%%%%%%%%%%%%%%%%%%%%%%%%%%%%%%%%%%%%%%%%%

\end{document}